\documentclass{article}

\usepackage{microtype}
\usepackage{graphicx}
\usepackage{subcaption}
\usepackage{booktabs} 

\usepackage{hyperref}

\usepackage[accepted]{icml2026}

\usepackage{amsmath}
\usepackage{amssymb}
\usepackage{mathtools}
\usepackage{amsthm}

\usepackage{xurl}

\usepackage[capitalize,noabbrev]{cleveref}

\theoremstyle{plain}

\theoremstyle{definition}

\theoremstyle{remark}

\icmltitlerunning{Irresponsible AI}

\begin{document}

\twocolumn[
\icmltitle{Irresponsible AI: big tech's influence on AI research and associated impacts}



  \icmlsetsymbol{equal}{*}

  \begin{icmlauthorlist}
    \icmlauthor{Alex Hernandez-Garcia}{mila,udm}
    \icmlauthor{Alexandra Volokhova}{mila,udm}
    \icmlauthor{Ezekiel Williams}{mila,udm,icl}
    \icmlauthor{Dounia Shaaban Kabakibo}{mila,udm}
    \icmlauthor{Mélisande Teng}{mila,udm}
  \end{icmlauthorlist}

  \icmlaffiliation{mila}{Mila, Québec AI Institute}
  \icmlaffiliation{udm}{Université de Montréal}
  \icmlaffiliation{icl}{Imperial College}

  \icmlcorrespondingauthor{}{alex.hernandez-garcia@mila.quebec, alexandra.volokhova@mila.quebec, e.williams1@imperial.ac.uk dounia.shaaban.kabakibo@umontreal.ca, tengmeli@mila.quebec}

  \icmlkeywords{Artificial intelligence, responsible AI, big tech, impacts}

  \vskip 0.3in
]



\printAffiliationsAndNotice{All authors have significantly contributed to this article. The order does not indicate amount of contribution.}  

\begin{abstract}
The accelerated development, deployment and adoption of artificial intelligence systems has been fuelled by the increasing presence of big tech in the AI field. This trend has been accompanied by growing ethical concerns and intensified societal and environmental impacts. This position paper argues that irresponsible AI development is strongly driven by big tech's influence and involvement in the field. First, we examine the growing and disproportionate influence of big tech in AI research and argue that its drive for scaling and general-purpose systems is fundamentally at odds with the responsible, ethical, and sustainable development of AI. Second, we review key current environmental and societal negative impacts of AI and trace their connections to big tech's influence. Third, we discuss the underlying economic forces driving big tech's actions. Finally, as a call to action, we invite AI researchers to counter big tech's influence in irresponsible AI development through strategies that build on the responsibility of implicated actors and collective action.
\end{abstract}

\section{Introduction}
\label{sec:intro}

In recent years, the technology known as artificial intelligence (AI) has shifted from being predominantly an academic study subject, to making recurrent headlines in mainstream media and becoming a conversation topic for many. While the vast majority of the world's population does not actively use AI or participate in its development \citep{statsintro}, AI sparks enthusiasm across a wide range of sectors for the opportunities it may offer. Meanwhile, the accelerated deployment and adoption of AI is responsible for tangible societal and environmental impacts \citep{crawford2021atlas,bender2025aicon}.

The transition of AI from academia into the public sphere has gone hand-in-hand with the corporate world, in particular ``big tech''. Here, we understand big tech as primarily the group of largest technology companies such as Google, Meta, Microsoft and Amazon, but our use of the concept pertains other AI-focused large companies, such as OpenAI. These companies play a central role in the public perception of AI and steer the agenda on many aspects of AI research, development, application, and even regulatory decision making for the field \citep{whittaker2021capture, jurowetzki2021privatizationai, ahmed2023industry, gizinski2024bigtech}. Importantly, big tech has dedicated efforts and resources to shape the narrative around responsible AI development, with many large corporate players writing their own responsible AI guidelines, engaging in AI ethics research and organising or sponsoring related events \citep{jobin2019global, young2022facct, bughin2025bigtechrai}.

Nonetheless, rather than instantiating responsible, ethical and sustainable practices when developing AI technologies, big corporations have significantly contributed to the negative global impacts of AI: environmental harm due to increased demands for energy and resources \citep{crawford2021atlas, desroches2025environmentalimpact}, increased surveillance \citep{zuboff2023surveillance}, loss of privacy \citep{veliz2021privacy}, infringement upon intellectual property \citep{jiang2023ai}, spread of mis/dis-information \citep{bontridder2021role, raman2024fake}, increased inequality \citep{adams2024new, kim2021ai}, degradation of labour rights \citep{altenried2020platform, crawford2021atlas, bender2025aicon}, etc. While the role of big tech in these negative impacts is often discussed in certain academic and non-academic circles \citep{abdalla2021bigtechbigtobacco, young2022facct, verdegem2022aicapitalism, bender2025aicon}, big tech's growing influence is often overlooked within the larger AI research community, likely due to its pervasive presence \citep{whittaker2021capture}.

\begin{figure*}[t]
\centering
\includegraphics[width=0.8\linewidth]{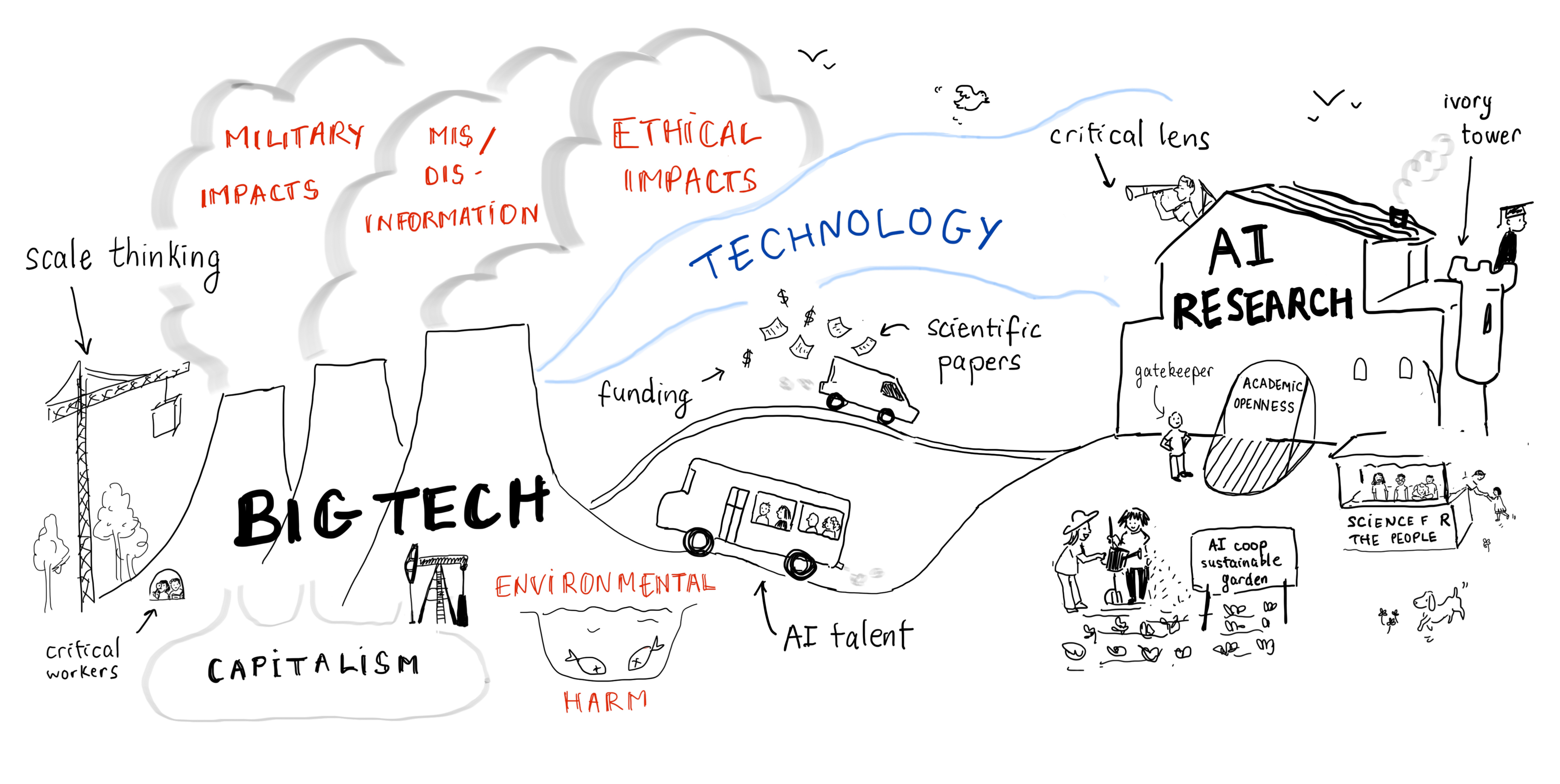}
\caption{An artistic summary of our position, depicting certain aspects of the influence of big tech in AI research, its associated impacts, and our call to action.}
  \label{fig:summary}
\end{figure*}

In this position paper, \textbf{we argue that big tech's influence on AI research is an important driver of irresponsible AI development}. We review literature from diverse fields with the aim of 1) summarising the influence that big tech has on AI research, 2) examining the link between the negative impacts of AI and big tech's influence, 3) discussing why big tech, and corporate tech more broadly, is incentivised and structured to favour irresponsible AI (iAI), and 4) suggesting ways in which researchers can counter this influence and thus support responsible AI efforts. Consequently, we echo the many voices asserting that technical solutions alone, without addressing factors such as corporate influence and industry incentives, will fail to result in truly responsible AI \citep{greene2019better, verdegem2022aicapitalism, kalluri2020don, adams2024new}. By shedding light on big tech's influence on irresponsible AI development, a topic often overlooked in AI technical circles, we present a lens through which we, as AI researchers, are invited to take a critical look at our field and to consider how our own research is situated in it. We hope to inspire our research community to take action and embrace the important role we can play in many potential solutions to iAI. 

\section{How does big tech influence AI?}
\label{sec:influence}


The technology industry is naturally interested in computer science research and has historically influenced its development \citep{mahoney1998history}. However, since the early 2010s, in what \citet{sevilla2022compute} named the Deep Learning and Large-Scale eras, the influence of big tech in artificial intelligence research has dramatically increased and is currently pervasive \citep{ahmed2023industry}. The modes of exerting this influence are multifarious \citep{bak2025risksinfluence}, and they are reflected in the research community in many ways.

\begin{figure*}[t]
\centering
\includegraphics[width=0.81\linewidth]{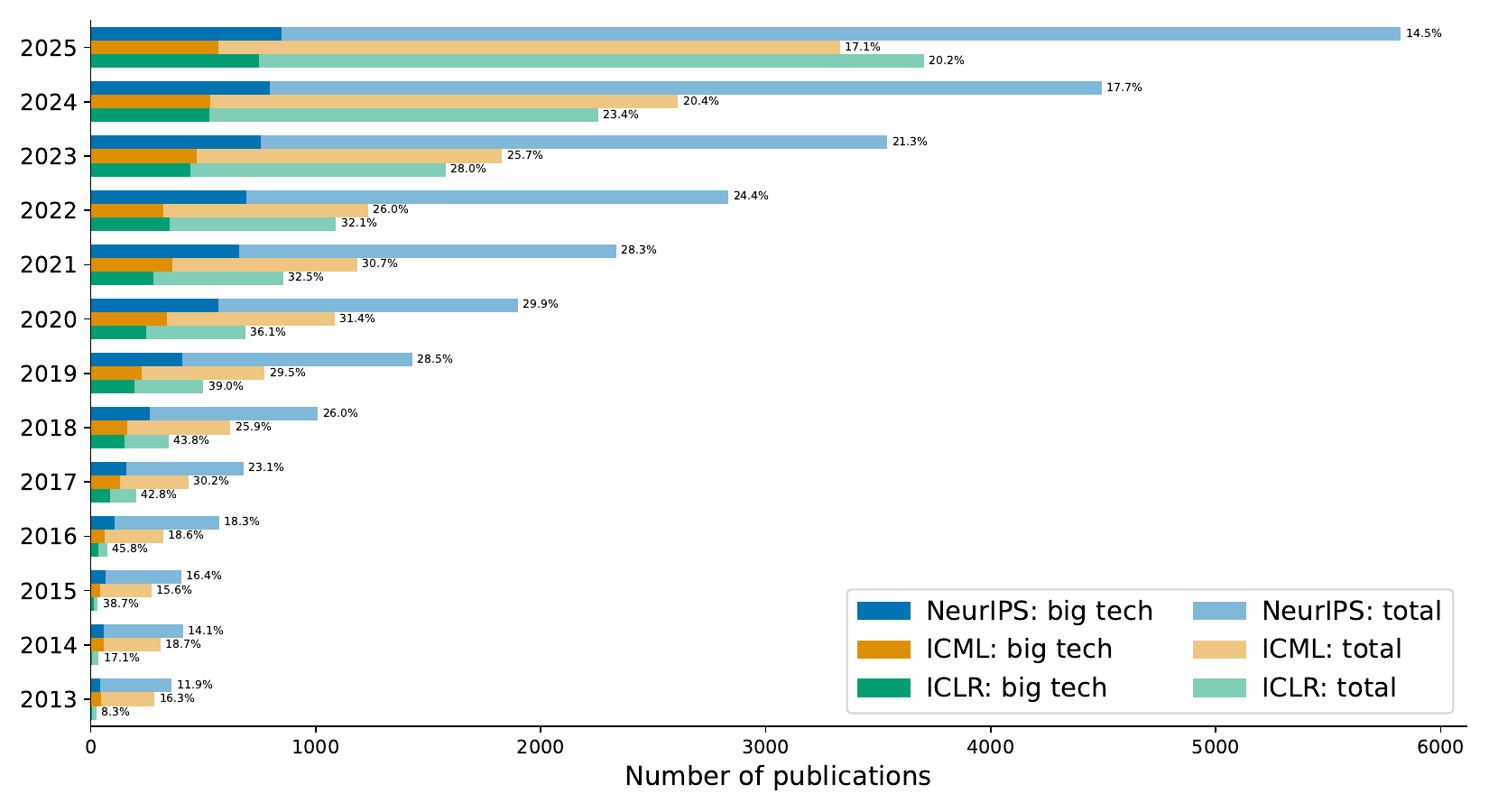}
\caption{Number of publications at NeurIPS, ICML, and ICLR in 2013--2025 and the fraction of them with big tech affiliations. For more details, see Appendix \ref{sec:app_pubs}.}
  \label{fig:bt_papers}
\end{figure*}

To assess the contribution of big tech to AI research, we have analysed the authorship in the publications at NeurIPS, ICML, and ICLR in the period 2013--2025 (\cref{fig:bt_papers} and \cref{sec:app_pubs}). The average percentage of the papers with at least one author affiliated with big tech constitutes 21.1\% at NeurIPS, 23.5~\% at ICML, and 31.4~\% at ICLR. Notably, the absolute number of big tech publications grows throughout the whole period, while the fraction of the total peaks in 2020 (NeurIPS and ICML) and in 2016 (ICLR) and decreases thereafter. The post-2020 decrease can be explained by the higher growth rate of the total publication number as well as big tech's transition towards less costly and time consuming non-peer-reviewed publishing media, as suggested by \citet{gnewuch2025}.

 In an independent study, \citet{ahmed2020dedemocratization} observed similar trends of industry presence across a larger set of conferences in 2000-2019. In addition, \citet{birhane2022values} observed that many of the industry papers are co-authored by researchers employed by the biggest players, such as Google and Microsoft---for instance, 167 publications at NeurIPS 2019 (more than 10~\% of a total of 1428 accepted papers) were co-authored by researchers affiliated with Google \citep{jurowetzki2021privatizationai}. Focusing on the most highly-cited papers at ICML and NeurIPS, \citet{birhane2022values} showed that the percentage of big tech affiliations increased from 13~\% in 2008/09 to 47~\% in 2018/19. This suggests increasingly strong impact on the AI research community. Notably, \citet{gnewuch2025} demonstrated that industry-funded papers have a preference for citing industry-funded publications, suggesting higher self-citation tendency, which contributes to maintaining prominence and moulding the agenda of the field for the interests of the industry. 

The presence of big tech in ML conferences goes beyond the accumulation of publications. For instance, big tech corporations consistently sponsor and thus fund the major machine learning conferences---NeurIPS, ICML and ICLR---along with many other AI-related events, including the conference on Fairness, Accountability, and Transparency (FAccT) \citep{young2022facct}. Furthermore, researchers with corporate affiliations occupy many of the major managing and organisational positions at these conferences. By way of illustration, 7 out of 12 of the board members of NeurIPS 2025, plus the Treasurer and the Secretary, were affiliated with corporate organisations \citep{neurips2025board}. Recently, ICML has announced a partnership with Google to incorporate one of their AI products for (optional) automated paper feedback \citep{kamath2026icmlgoogle}. Finally, big tech is also behind the organisation of many workshops at the major conferences and many invited speakers are industry researchers.

Another example of the prominence of big tech in AI research is the extraction of research talent from academia to industry. According to data from \citet{ahmed2023industry}, the number of PhD graduates from AI-related fields in US and Canadian universities that went to industry increased from 21~\% in 2004 to 70~\% in 2020. Regarding faculty, the number of professors who transitioned from academia to industry has increased eightfold since 2006. Joint faculty appointments between universities and the industry have also increased, and these trends are having consequences in terms of department culture shift, research directions, conflicts of interest and the quality of the mentorship students receive \citep{morrisett2019academiaindustry, whittaker2021capture}. 

A widely consolidated practice that illustrates the permeation of big tech into academia is the widespread adoption of their software by universities. Despite the existence of free and open-source software, researchers and workers are often required to non-consensually use email services, office software and data management tools (to name a few) from Google and Microsoft. Recently, these tools have been updated with AI-powered chatbots and functionality and many universities are uncritically adopting and integrating generative AI into their standard software and procedures \citep{guest2025aiacademia}. This renders researchers and workers more dependent on big tech software and makes them look less critically at the potential negative impacts.

While further detailing the influence of big tech in AI research is beyond the scope of this paper,\footnote{A comprehensive analysis of the implications of big tech in academia, and a comparison with the well studied influence of Big Tobacco, was offered by \citet{abdalla2021bigtechbigtobacco}. \citet{gizinski2024bigtech} offer a recent review of the literature on big tech's influence on AI research and study how its reach and power impact the propagation of ideas. \citet{bak2025risksinfluence} recently analysed the risks of the tech industry on research, public health, politics and the natural world, as well as a description of the mechanisms through which this influence is exerted.} a consequence worth highlighting for our purposes is the industry's capacity to influence the research agenda of the field \citep{bak2025risksinfluence}. While big tech's research interests are broad, we here identify and highlight two relevant and related threads: the scaling paradigm \citep{hanna2020againstscale} and the push for general-purpose systems \citep{verdegem2022aicapitalism, bender2025aicon}.

Scale thinking \citep{hanna2020againstscale, varoquaux2025hype} refers to the belief that the main driver for progress in AI is increasing the amount of training data, the size of the models, etc. without necessarily innovating on the fundamental aspects of the technology. This belief was particularly popularised in the ML community by \citet{sutton2019bitter} and his bitter lesson that general-purpose models that scale with computation ultimately perform better than methods relying on human knowledge; and by an OpenAI paper which proposed that neural language models performance improves predictably as amount of training data, compute and number of parameters are scaled \citep{kaplan2020scaling}. 
It is then not surprising that large corporations follow a scaling approach in their research endeavours, given their access to enormous amounts of data and compute power, which gives them a competitive advantage relative to academia. Furthermore, the biggest corporations have a business interest in establishing this trend, since cloud computing and data centres is an important part of their business model---Amazon, Google and Microsoft alone control more than two thirds of the cloud computing market \citep{widder2025big}.

\citet{luccioni2023countingcarbon} discussed the low correlation between model performance and the energy (and carbon emissions) used for training. More recently, \citet{varoquaux2025hype} have discussed the scientific fragility of the scale thinking paradigm in connection to the diminishing returns of model size, arguing that an appropriate model architecture plays a bigger role in increasing performance and that in many applications smaller models perform equivalently or better than larger counterparts.
While scaling may be a legitimate approach in certain cases, it has been an outsized focus of big tech, fuelled by the dominant growth imperative (\cref{sec:capitalism}), and, as we review in \cref{sec:consequences}, is at the core of much of irresponsible AI development.

Similar to the scaling approach, another prominent theme in big tech's research is the drive towards general-purpose methods, epitomised by the Artificial General Intelligence (AGI) slogan.
The pursuit of general-purpose systems is also reflected in current commercial chatbots, which are designed and sold as usable for a wide range of tasks. Like the scaling paradigm, this is also tied to big tech's business model and competition for monopolising and controlling the technology market: a versatile chatbot allows the system provider to offer a range of services which could otherwise be provided by multiple companies. The design philosophy behind this approach is that AI systems should be task-agnostic, as opposed to tailored and domain-inspired approaches. For example, in the application of ML for chemistry, one approach is to adapt the methods to the specific application---small molecules, proteins, DNA, inorganic crystals, amorphous materials, etc.---via careful data selection and domain-inspired constraints \citep{krenn2022selfies,milaforscience2023crystalgfn,therrien2026obelix,cretu2025synflownet}; the general-purpose approach typical of big tech pursues broad tasks with maximal data and compute, as exemplified by the recent paper by Meta researchers in which a very large diffusion model was trained jointly on data sets of both molecules and materials \citep{joshi2025allatom}. General-purpose and scaling might improve the performance on certain benchmarks and bring about practical benefits. However, it is reasonable to question the cost of these paradigms, and whether alternative approaches can bring the same or even higher benefits. We discuss how general-purpose systems lead to irresponsible development in the following section.

\section{How does big tech contribute to irresponsible AI?}
\label{sec:consequences}

The push for larger and general-purpose systems by big tech comes with a social and environmental footprint \citep{bhardwaj2025limits,uuk2025generalpurpose}. A now well-studied impact of this approach are the carbon emissions of large-scale training and deployment of AI systems \citep{luccioni2023countingcarbon, luccioni2024inference, varoquaux2025hype}. AI models, especially those from large corporations, are typically trained and deployed in large data centres which consume enormous amounts of energy \citep{devries2023energy, guidi2024environmental}, fresh water \citep{li2025water, barratt2024revealed} and raw materials \citep{crawford2021atlas}.

The AI infrastructure is mainly built and controlled by big tech: Amazon, Microsoft, and Google account for 63~\% of the global cloud market as of April 2026, when it reached a 35~\% year-on-year growth rate---its highest rate since 2022 \citep{synergy2026}, closely tracking the generative AI boom. This suggests that environmental costs are not incidental to big tech's AI agenda but constitutive of it, and concentrated within precisely the actors driving the scaling agenda. This is epitomised by massive ventures like Meta’s largest data centre in Louisiana, which could consume up to 30~\% of the state’s power supply, and Project Stargate---a joint venture by OpenAI, Oracle, and SoftBank, consisting of a \$500 billion data centre that would be powered by a \$500 million natural gas plant---as well as the \$10 billion data centre investments by Microsoft, Google and Amazon in Mexico, which will require large quantities of fresh water in regions increasingly prone to droughts \citep{baptista2024thirsty, thomas2024mexico}. As a result, the reported carbon emissions by big tech and their projections have soared \citep{milmo2024emissions} and they have dropped their already weak emission compensation pledges \citep{hoffmann2022pledges, marx2024climatedisaster}. This suggests that corporate climate and responsible AI pledges are often business tactics rather than signals of proactively striving to go beyond mere compliance to regulatory frameworks \citep{foroohar2019dontbeevil}. All this has led researchers to question whether AI, and even computing more generally, could ever be sustainable \citep{becker2023insolvent, schutze2024sustainableai, rehak2025sustainableai}.

Besides backtracking on climate promises, big tech often abandons ethical considerations too, as a result of subjecting to the imperative of scale and growth.
Many scholars have studied how the development and application of AI at scale has often resulted in reinforced inequality, discrimination and exploitation of workers \citep{oneil2017weapons, west2019discriminating, bender2021parrots, benjamin2023race}. As shown by \citet{widder2023dislocated}, AI practitioners tend to disregard ethical work when it is perceived as an obstacle towards scaling of their systems, which in turn is seen as an unquestionable merit. The scaling paradigm reinforces the distance between the developers of a technology and the people affected by its use and development. This distance ``makes it possible to not see harm'' and facilitates a modularised, transactional attitude towards AI development, where no social relationship is created when a piece of technology is deployed. For example, this distance plays out in the practice observed across big tech companies of utilising text and image datasets composed of literature and art created in large part by artists who rely on their intellectual property rights to make their living, without remunerating them \cite{jiang2023ai, bender2025aicon, alex2025searchlibgen}. 
The appropriation of the work of artists to train AI models poses a threat to their jobs, which provide significant satisfaction and purpose in addition to artists' livelihoods, beyond the use of unlicensed copyrighted work. \cite{jiang2023ai, paris2025tech}. Indeed, the generative AI models that are built from artists' original work are increasingly used in creative industries such as illustration and graphic design. Artists have raised their voices against the proliferation of AI generated artistic content, for example with the $\#$StarterPackNoAI digital protest movement of illustrators, and the launch of the \textit{Is This What We Want?} silent album which brought together over 1,000 artists \citep{isthiswhatwewant}, and whose tracks titles form the sentence ``The British Government must not legalise music theft to benefit AI companies''.

A growing concern with impacts that are still uncertain is the proliferation of so-called AI slop due to the widespread adoption of generative AI \citep{holden2026aislop}. According to a study by \citet{paredes2026genai}, about half of all internet articles in English are now primarily generated by language models. While some authors have discussed the cultural value of AI slop \cite{kommers2025aislop}, others speak of AI as trash \citep{pasek2025aitrash} to highlight both the informational and environmental harms. In fact, the generative AI systems deployed by big tech have been connected to the spread of mis- and dis-information \citep{bontridder2021role, wack2025generative}, and social media platforms owned by big tech have played a direct role in the dissemination of fake news \citep{aimeur2023fakenews}. In academia, there are growing concerns that the massive, uncritical adoption of generative AI to write papers and peer reviews will have a negative impact on the quality of science \citep{liang2025llms, kim2025position, guest2025aiacademia}.

Big tech’s impacts also extend into militarism, where AI and data resources are deployed to support the scaling and proliferation of warfare \citep{ali2021artificial, maaser2022bigtechwar}. For instance, Google, Amazon, and Microsoft provide cloud services, AI tools, and infrastructure to the Israeli military \citep{ohchr2025from, fatafta2025artificial, abraham2024order, davies2025revealed}, which has killed tens of thousands of Palestinians in Gaza since October 2023.
Since late 2023, South Africa has an open proceeding against the State of Israel at the International Court of Justice concerning alleged violations in the Gaza Strip of the Convention on the Prevention and Punishment of the Crime of Genocide \citep{icj2024genocide}. In September 2025, a UN Independent International Commission concluded that the actions committed by Israeli authorities against Palestinians constitute ``underlying acts of genocide'' \citep{hrc2025genocide}.
Google has recently announced a partnership with Lockheed Martin---a major weapons contractor---to ``integrate Google's advanced generative AI into Lockheed Martin’s [...] ecosystem" \citep{lockheed2025lockheed}. Previously, Meta also announced partnerships with other surveillance and weapons companies such as Palantir and Anduril \citep{clegg2025open}. Amazon, Microsoft, and Google have all contributed to Project Maven \citep{manson2024ai}, a US military initiative utilising AI to assist decision-making in military operations. 

Contrary to common claims that AI will make warfare more precise and reduce bloodshed, AI use has reportedly played a key role in Israel's levelling of Gaza, with algorithms being used to facilitate and justify high-throughput targeting and striking \citep{abraham2024lavander}. Military and geopolitical harms are exacerbated by large, general-purpose systems used across arbitrary domains, since they are inherently dual-use, as oppose to narrowly purposed systems. Associated harms are compounded by the geopolitical pressure to control these systems before adversaries do---a classic arms race dynamic. Moreover, since geopolitical conflicts increase the demand for weapons, profit-driven weapon manufacturers have a long history of lobbying for the proliferation of wars \citep{hartung2004much}. We speculate that the growing involvement of big tech and AI technology in this business amplifies pro-military political influence and increases the economic incentive for expanding conflicts around the globe \citep{volokhova2026implicated}.



\section{What influences big tech?}
\label{sec:capitalism}

To draw a more general picture, this section explores what influences big tech to engage in iAI. We argue that it is primarily pressure for economic growth and profit, and discuss reasons why government regulation appears to be ineffective at curbing iAI. While such underlying forces influence all corporations, the nature of big tech and the AI industry may exaggerate their consequences. We discuss economic growth seeking, followed by governmental regulation, next.

Because big tech is composed primarily of for-profit, capitalist firms, their decision making is most often driven by a \emph{growth imperative} \citep{richters2019growth, verdegem2022aicapitalism}. This is the idea that an entity, beyond simply seeking enough profit to sustain itself, is pressured to grow. While there is still debate on the subject \cite{lawn2011steady, richters2019growth}, theoretical and empirical evidence over the last century suggests that such imperatives are a reality, particularly for corporations \cite{marx1867capital, meadows1972thelimits, ferguson2018growth, hickel2020less}. It is believed that growth imperatives stem from forces present in the capitalist economy, for example, from the need for entities to compete for market share and new investments \cite{gordon2003capitalism, richters2019growth}. Finally, past literature highlights the role of science and technology innovations in creating additional growth pressure: to survive, a company must grow so that it can re-invest in research and development, leading to product or operating advances, to keep up with the innovations of competing companies \citep{cefis2005matter, schumpeter2013capitalism, richters2019growth}. This constant pressure to innovate might partly explain the strong implication of big tech in AI research discussed in \cref{sec:influence} \citep{rikap2022bigtech}.


We speculate that, via AI investment, big tech endures a growth imperative quite acutely. Currently, many AI products remain unprofitable \cite{kost2025ai} and therefore certain companies appear to rely on the perception that they will keep ``growing'' to obtain further investment and avoid bankruptcy \cite{bender2025aicon, doctorow2025pluralistic}. While not conclusive evidence of this, the fact that AI companies are one of the biggest drivers of US market growth \cite{hyatt2025ai, aliaga2025is} is consistent with the idea. Though most big tech companies remain highly profitable, they are intertwined with the AI industry not only via in-house research but by heavy investment in AI firms \cite{widder2025big}. Thus, the current unprofitability of AI coupled with AI innovation---in product, process, or even just perception---may drive heightened growth pressure not just for smaller AI companies and start-ups but for big tech.

In light of the economic evidence, it seems that profit and growth seeking disproportionally influence AI. But do growth imperatives particularly drive the irresponsible impacts of big tech? We speculate that such a link exists, based on historic examples outside and inside tech. Outside tech, the oil industry has notoriously misled the public around the impacts of climate change in a bid for continued profit \cite{dembicki2022petroleum}. Similar patterns with dire consequences have been well studied regarding the tobacco and other industries \cite{brownell2009tobacco}. Uber provides an example inside tech. The company originally pursued aggressive global expansion fuelled by venture capital, and mainly operated at a yearly loss between 2014 and 2022 and only recovered profitability in 2023 \citep{curryUberStats2025}. Throughout this period, Uber implemented precarious gig work, brought about the collapse of traditional taxi livelihoods, and increased urban congestion, all while frequently ignoring trade regulations \citep{smith2018uber}.

To balance unrestrained profit and growth seeking, liberal thinkers often look to government regulation. However, the evidence from sections \ref{sec:influence} and \ref{sec:consequences} suggest that, thus far, regulation has done little to rein in iAI. In certain cases, governments have even been unable to properly tax big tech firms \citep{zafiroski2023taxbigtech,canada2025canada}. We see two reasons for this paucity of regulation. First, big tech is able to leverage the wealth and power acquired in their pursuit of profit, coupled with what economists describe as the \emph{monopolization of knowledge}, \citep{pagano2014intellectualmonopoly, widder2025big}, to escape regulation \citep{rikap2022bigtech} that might otherwise limit iAI. Specifically, power and wealth can be leveraged for lobbying and funding politicians: in 2025, the top AI and big tech firms spent nearly \$250 million on US government lobbying \cite{techoversight2025lobbying}. Second, governments themselves face growth imperatives in the face of a primarily capitalist global economy, and aim to maintain growth to keep measures like currency value and unemployment stable \cite{ferguson2018growth, richters2019growth}. This can lead them to under-regulate industries that are viewed as strong drivers of growth---a dynamic that seems highly likely with AI. Government press releases advocating for accelerated AI adoption in the name of growth provide evidence for this \cite{department2025ai, european2025commission}. In a press release on its AI strategy task force, the government of Canada states its commitment to building an economy ``that prioritizes innovation and the use of emerging and digital technologies to support growth." \cite{canada2025government}.

In summary, we emphasize that iAI can be traced to fundamental features of capitalism---such as capital accumulation and competition \cite{harris2023capitalism}---and can exacerbate problems like run-away growth pressure and corporate capture. Thus, technical solutions alone cannot fully address irresponsible AI. AI researchers might thus conclude that iAI falls outside of their purview, or is too overwhelming a problem to consider. The next section outlines multiple paths to addressing it that rely on the research community, so we encourage the reader to not lose hope.

\section{Call to action: How can we mitigate big tech's irresponsible influence?}
\label{sec:solutions}

To answer the question of what we, as AI researchers, can do about the societal and environmental harms caused by AI under big tech's influence, we suggest to first look into how we are situated in this context. It is important to acknowledge that due to corporate involvement and AI popularity, our field is financially privileged, with funding opportunities and higher salaries \citep{Harroch2026AICompensation}. Especially in the Global North, those who can travel to attend prestigious conferences around the globe enjoy free food, luxury parties and other entertainments offered by corporate sponsors. With that, we find ourselves in a position of \emph{implicated subjects} \citep{rothberg2019implicated} with respect to the harms described in \cref{sec:consequences}, since we are indirectly benefiting from the system creating them. This position of implication can be uncomfortable to recognise, but it need not result in feelings of guilt. Instead, it invites us to critically reflect upon our own situation, which brings opportunities for transformative action. Change happens in complex, non-linear ways and there are often multiple avenues for effective contributions ranging from individual to collective action and depending on personal circumstances \citep{duncan2016change}.

Starting with our own research projects and studies, we recommend critically examining how corporate incentives---through funding, publications, or collaborations with industry---may shape the research questions posed and the methodologies employed. Assessing who stands to benefit from the work, and who may be disadvantaged, is essential for uncovering these influences in both theoretical and applied research. Researchers involved in event organisation may consider alternative funding options to big tech, or require that sponsors avoid research with applications in domains with a high risk for iAI, like weapons or surveillance technologies. Relevant literature exploring the intersections of profit and power in AI provides valuable guidance for such critical reflections \citep{ali2021artificial, kalluri2020don, adams2024new, bender2025aicon}. Papers and books which we choose to read and discuss with our colleagues shape both our imagination and our future research directions, as well as those of our peers. Creating spaces for critical discussions, for example, a reading group, is a valuable step towards rethinking and redirecting research endeavours.

Echoing the work of critical scholars \citep{hanna2020againstscale, mohamed2020decolonial, bender2021parrots, widder2023dislocated, carroll2023care}, we suggest centring the needs of communities and the building of social relations in the research and development of new technologies. Concretely, this means informing projects by the specific needs of community groups, and ideally involving them in their design. For example, projects on geospatial data modelling for agriculture should be informed by the needs of farmers. This approach connects with the principle of developing situated, narrowly scoped technologies supported by ethical labour and data practices \citep{bender2025aicon, carroll2023care}. Ethically curated, community-centred research has the potential to strengthen social connections and empower communities, as opposed to research advancing large, general-purpose AI systems which may exacerbate social inequalities and the climate crisis for the benefit of a few big tech shareholders.

In order to conduct research grounded in these ethical principles, one needs to have a sufficient level of control over the project, which individual AI practitioners often lack \citep{widder2023s}. While individual actions alone may not be sufficient to shift industry incentives, these practices are key to developing the critical consciousness that transforms individual researchers into actors capable of participating in and eventually leading institutional reforms, as discussed by \citet{thompson2025cv}, and lead towards building collective (counter)power that can play a crucial role in mitigating big tech negative influence. An example of a campaign that started as a small-scale initiative and eventually raised awareness among the general public is the recent \citet{enabledemissions}, that helped expose the partnerships between big tech and the fossil fuel industry.

Another exciting path to countering the power amassed by big tech, and the economic forces that drive it toward iAI, is to use alternative forms of entrepreneurship to build new organisations that are more democratically controlled or that are not as influenced by the growth imperative, such as cooperatives and non-profits. Contributing to worker co-ops instead of venture capitalist-funded start-ups can avoid wealth and power centralization via more democratic business structure, and be readily organised around fulfilment of societal needs rather than investor-driven pressure to grow \cite{verdegem2022aicapitalism}. There is abundant precedent for co-ops outside of tech such as in agriculture and electrification which have operated at scale for years, demonstrating long-term economic viability. The relative scarcity of cooperative structures in AI does not imply that they are unviable; rather, it reflects path dependencies and cultural norms in the field, currently heavily shaped by major for-profit players. However, we note a growing interest and practice in cooperative models within tech and AI \citep{scholz2025coopai}, including academic literature and concrete business examples that researchers interested in co-ops as a lever for change might use for inspiration. Scholars have proposed worker-owned cooperative models for AI training data \citep{sriraman2017coops}. Digital platform cooperatives governed by their users or workers \citep{zhu2021coops} and non-profits offer a democratic alternative to corporate platforms that privilege shareholder wealth maximization. \footnote{Examples of tech co-ops include Hypha (Canada), which supports co-op infrastructure, decentralized social media, and privacy-preserving AI; Baseline (Canada), a co-op that offers AI and software development support services; Loomio (New Zealand) develops open-source communication software for democratic decision-making; and Transkribus (Europe), an AI system for recognising and transcribing historical and handwritten documents. Similarly, tech non-profits pursuing related goals include Codeberg (Germany), which provides web services for open-source projects, and Lanfrica Labs (Canada), which organises African AI resources from datasets to papers and models.} 



Researchers who are not situated to start a business, but would like to pursue collective action in their free time, can join relevant grass-roots organisations or political movements. An example grass-roots organisation is Science for the People \citep{sftp2025} (SftP). SftP was founded by MIT scientists who opposed the use of their work in the Vietnam war, and has since expanded beyond the US, with various chapters around the world. It follows a centralized approach to science communication---with a periodical magazine focused on specific topics---and a decentralized approach to local organising, where SftP chapters work on problems relevant to their local communities. Tech Workers Coalition \citep{twc2025}, and No Tech for Apartheid \citep{ntfa2025} represent other examples. Getting involved with organisations like these can yield insight into adopting a more community-oriented research paradigm, and can leverage their size to take on projects and apply political pressure in ways that individuals cannot. For example, activist movements have often stood out as the catalyst of massive institutional changes, such as the civil rights movement, suffrage, and the Indian revolution. 

Depending on one's domestic political landscape, there may also be opportunities to volunteer with political parties at the municipal, regional, or state level that actively campaign to regulate big tech, or invest in publicly owned tech infrastructure that is less impacted by economic pressure. Alternatively, community organisations and workers co-ops may serve as bases for organising to lobby political parties to counteract irresponsible outcomes of growth pressure. 

For researchers in big tech and industry, there are many valuable ways to work towards responsible AI ``from the inside''. The specific actions may vary substantially based on the researcher's position in a company. An accessible action for many could be simply raising awareness among colleagues, and fostering an understanding not just of positive, but negative, impacts of their workplace's research practices. In positions closer to leadership, workers might be able to advocate for the development of more bespoke models, for specific problems, and a decreased pursuit of general models and scale. Depending on circumstances, collective actions like petitions, walk-outs, and sit-ins, or whistle-blowing, can build public awareness and political pressure that can be harnessed to correct irresponsible practices \citep{pbs2018maven, king2019lucas, mccanne2022when, koren2022google, grant2022google}. Perhaps most crucially, forming a union represents a great step towards responsible AI, as unions can be used as a vehicle for broader system change, in addition to workers' rights \citep{lewin2024emerging, whittaker2021capture}. For example, unions have been essential in winning many of the society-scale labour rights that we take for granted today in many parts of the world, including weekend breaks, hour limits, parental leaves, safety standards, and the ban of child labour.

\section{Alternative Views}
\label{sec: alternatives}

\paragraph{Big tech's influence on the research agenda enables progress}
In this paper, we have argued that big tech's outsized influence in steering the AI research agenda towards scaling and the pursuit of general-purpose AI systems can lead to irresponsible AI. A common counterargument is that pursuing this research agenda has acted to create exciting AI models \citep{brown2020language, touvron2023llama, team2023gemini, jumper2021highly, kirillov2023segment} that have driven significant scientific, technological and economic progress, with projections suggesting meaningful global GDP growth over the coming years \citep{goldman2023generativeai}. Moreover, the development of general-purpose AI systems could be justified by analogy with earlier general-purpose technologies such as electrification or the computer, both of which generated economic gains. As a counterargument, we challenge the underlying assumption that the supposed progress outweighs the current and potential socioecological impacts of the rapid deployment of AI systems at scale (see \cref{sec:consequences}). Furthermore, we argue that it is questionable that the recent AI development and the potential scenarios aimed at by big tech would truly benefit the general population \citep{doctorow2025pluralistic}. Finally, we note that the effects of AI may emerge more rapidly than those of earlier widely adopted technologies, since much of the necessary digital infrastructure is already in place and these systems are relatively easy to adopt, often requiring only natural language interaction \citep{mcafee2024generallyfaster}. This raises heightened concerns about societies’ ability to anticipate, govern, and mitigate negative impacts.

\paragraph{Environmental impacts of big tech are overstated}
Some argue that the impacts from the rapid deployment of data centres are overstated in aggregate terms, noting that data centre consumption as a share of global electricity demand increased from 1~\% in 2005 to 1.5~\% in 2024 \citep{IEA2025}. We highlight two problems with this argument. First, as noted by \citet{masanet2024better}, many uncertainties stem from limited clarity in reporting on current data centre energy use and operating conditions, which complicates both present analysis and future projections. Second, aggregate measures can obscure more salient localised effects. The environmental burden of AI infrastructure is highly dependent on where data centres are built, how local grids are stressed, and which communities bear water, land-use, and heat externalities.

 For example, in regions such as the Canadian province of Quebec, electricity generation is predominantly renewable \citep{CER_Quebec_Energy_Profile} and tightly regulated, resulting in a comparatively low energy carbon footprint for AI infrastructure. In contrast, energy production in the United States---which accounted for the largest share of global data centre electricity consumption in 2024 (45~\%) \citep{IEA2025}---remains heavily reliant on fossil fuels \citep{eia2025monthlyenergyreview}. Privatized utility structures in the United States have also been shown to create incentives that discourage building at scale infrastructure for green energy \citep{lusiani2024entrenched}. In this context, additional electricity demand from AI development risks reinforcing existing carbon-intensive pathways rather than accelerating a transition toward cleaner energy. Similarly, water--abundant regions face less concern over data centres’ consumption, but many are being built in water-stressed areas \citep{baptista2024thirsty}. In 2023, Microsoft reported that 42~\% of its water consumption came from areas with ``water stress'' \citep{microsoft2024env}, while Google stated that 15~\% of its freshwater withdrawal occurred in regions with ``high water scarcity'' \citep{google2024env}.

\paragraph{Top-down structural reform should be the priority}
Many of the solutions proposed in \cref{sec:solutions} require involvement and initiative from individual researchers, or groups of researchers, with the aim to empower them. However, some may argue that such approaches risk placing a disproportionate burden on individuals who might have limited institutional power. From this perspective, responsibility for meaningful change should not rest primarily on individual actors and collective action, but rather on policy-makers. In other words, it is top-down leadership that should establish guardrails on corporate behaviour. Examples include the (revised) Energy Efficiency Directive in the European Union \citep{eu2023eed} which requires data centres to systematically track their energy consumption and to evaluate and disclose their energy performance. However, the trajectory of this directive also illustrates the limits of top–down reform in isolation. The 2024 delegated regulation \citep{eu2024dr1364} that builds on the previous Directive introduced a confidentiality clause keeping individual data centre metrics out of public view, with language that, according to an investigation by Investigate Europe \citep{joyner2026meps}, closely mirrored proposals submitted by Microsoft and DigitalEurope.
We agree that top-down approaches can play an important role in tackling iAI. However, as this example illustrates, regulatory guardrails can themselves be shaped by the same pressures they aim to constrain. Moreover, there appears to be meagre historical precedent for top-down changes occurring autonomously. That is, this alternative view begs the question of how top-down change could occur in the absence of bottom-up pressure from researchers and other stakeholders, especially in light of the economic incentives that corporate and government leaders face (see \cref{sec:capitalism}).

\paragraph{Capitalism as the most viable current option}
In \cref{sec:capitalism}, we have argued that the dominant strategies of big tech in AI research and development are not incidental, but are deeply shaped by the logics of the current economic system. Some may contend that the contemporary capitalist model is ``what works best'', despite the shortcomings. There are differing views on how these shortcomings might be mitigated, from keeping the current status quo but imposing tighter regulations, to more fundamental transformations of the economic system. We leave this question open, inviting the reader to reflect on the broader structural conditions that shape the research and development of AI.

\paragraph{Alternative actors could have led to similar outcomes}
One might argue that irresponsible AI development could still occur even in the absence of big tech actors. While this is a valid hypothesis, we argue that the unique role played by big tech---notably in the push for scale and general purpose systems---creates conditions and particularly strong incentives toward irresponsible development. Whether alternative configurations of actors and incentives could have led to similar outcomes lies beyond the scope of this paper, but remains an important question to keep in mind. 

\section{Conclusion}
\label{sec: conclusion}

Computer science and machine learning research may be able to provide technology and tools to improve human well-being. However, the current trends in AI are leaving societal and environmental footprints that seem at times to overshadow the potential benefits. In this paper, we have reviewed literature arguing that the disproportionate influence of big tech in AI research, development, and deployment significantly contributes to these harms and negative impacts. Our review then highlights, in turn, how big tech's accumulation of wealth and power, and its use of these to influence AI, may indeed be traced to features of the capitalist economy. While we focused on the largest of the big tech companies, the mechanisms of influence on AI research and their associated impacts we described can also apply to smaller industrial actors---whose priorities may arguably cascade down from big tech---which, in turn, can influence AI research, albeit at a smaller scale. Therefore, while acknowledging the major role played by big tech in irresponsible AI development, we invite AI researchers to an exercise of critical thinking that goes beyond categorising companies into big tech or not, in order to learn to recognise irresponsible AI practices, and mechanisms leading to it in contexts where they may not be obvious. 
We hope the strategies presented in \cref{sec:solutions} might inspire our colleagues to join us in taking steps to mitigate big tech's influence on irresponsible AI development. Science and technology developed both by and for the people have the potential of equitably serving us all. 

\section*{Acknowledgements}

We are grateful to the co-organisers and participants of the Critical Science reading group at Mila for stimulating discussions that helped shape this work. We also thank Vincent Mai for valuable feedback. Finally, we are grateful to all the scholars, journalists and activists whose work has inspired and educated us. Dounia Shaaban Kabakibo and Ezekiel Williams acknowledge support from the FRQ Nature and technologies Sector – Doctoral Research Scholarship Program. Alex Hernandez-Garcia acknowledges funding from IVADO and the Canada First Research Excellence Fund. Alexandra Volokhova acknowledges funding from CIFAR which supported her work on this project.

\section*{Impact Statement}

This paper examines the disproportionate influence of big tech on AI research and development, and its associated societal and environmental impacts. While our focus is on a handful of large corporations, one might observe that this big tech ``mindset''---centred on scale and general purpose systems---is spreading well beyond them, permeating small companies, and shaping the perspectives of young researchers navigating careers in academia, industry, or entrepreneurship. We hope that this paper encourages readers to critically reflect on this prevailing way of approaching AI, to recognize how they are situated within these structures, and to consider alternative paths forward. Finally, while primarily addressed to the AI research community, much of this paper is accessible to a general audience, and we hope it contributes to broader public awareness of the structural forces shaping AI today.


\bibliography{references}
\bibliographystyle{icml2026}


\newpage
\appendix
\onecolumn

\section{Analysis on the publications at NeurIPS, ICML, and ICLR}
\label{sec:app_pubs}

We analyse affiliations of the publications at NeurIPS, ICML, and ICLR in the period 2013--2025 as outlined in Section \ref{sec:influence}. For this analysis we downloaded the papers from the official proceedings websites (\url{https://papers.nips.cc}, \url{https://proceedings.mlr.press}) and from OpenReview (\url{https://openreview.net/group?id=ICLR.cc}) and automated the identification of affiliations using the \texttt{pymupdf} package and regular expressions. This allowed us to get affiliations for the majority of the papers, except for around 1.3-3.2~\% for each year, which we were not able to process due to unconventional formatting or to lack of affiliations information in the PDF file. 

Then, we search for the keywords associated with big tech in the affiliations to identify the papers with at least one industry-affiliated author. We defined the list of big tech companies based on the one suggested by \citet{birhane2022values} with the following keywords:  

\texttt{Alibaba}, \texttt{Amazon},  \texttt{AWS}, \texttt{Apple}, \texttt{Element AI}, \texttt{ServiceNow}, \texttt{Facebook}, \texttt{Meta}, \texttt{Google}, \texttt{DeepMind}, \texttt{Huawei}, \texttt{IBM}, \texttt{Intel}, \texttt{Microsoft}, \texttt{Nvidia}, \texttt{OpenAI}, \texttt{Samsung}, \texttt{Baidu}, \texttt{Salesforce}, \texttt{Tesla}, \texttt{Uber}, \texttt{Anthropic}, \texttt{xAI}

Note that we extended the list of \citet{birhane2022values} by adding AWS as a variation for Amazon; ServiceNow as the company which acquired Element AI; and Meta as a rebranding of Facebook. Moreover, we included additional companies such as Baidu, Salesforce, Tesla, Uber, Anthropic, xAI.

The code developed to carry out this analysis is available on the repository \url{https://github.com/AlexandraVolokhova/irresponsible-ai}.

\end{document}